# Pairing interaction from Demons in $Sr_2RuO_4$

Young Woo Choi[1], Jisoon Ihm[2] and Marvin L. Cohen[1]*

[1]*Department of Physics, University of California, Berkeley, CA 94720, USA and Materials Sciences Division, Lawrence Berkeley National Laboratory, Berkeley, California 94720, USA*

[2]*Department of Semiconductor Physics and Engineering, University of Ulsan, Ulsan, Korea*

**Abstract**

We investigate the properties of the recently observed "demon" mode, a 3D acoustic plasmon, in $Sr_2RuO_4$ with an emphasis on evaluating its role for the pairing interactions in this superconductor. The demon mode is a low-energy electronic excitation, and it has been suggested that it could contribute to a reduced Coulomb repulsion and even a possible attractive interaction between electrons. In this study, we explicitly calculate the dynamically screened Coulomb interaction for $Sr_2RuO_4$ by using a renormalized tight-binding band structure and the random phase approximation for the dielectric function. Although the focus here is on $Sr_2RuO_4$, this material is considered mainly as a prototype system, having an observed demon mode, and our results should be considered as a guide for application to other systems. Our calculations show that there are regions in ($\mathbf{q}$, $\omega$) space where the Coulomb interaction becomes attractive. We find that, although the demon mode is not capable of producing a total attractive electron pairing interaction in $Sr_2RuO_4$, it does contribute to a significant reduction in the Coulomb repulsion at the relevant pairing energy scale.

**INTRODUCTION**

Recently, an experimental observation of a demon-like mode has been reported in an unconventional superconductor $Sr_2RuO_4$ [1]. This important discovery raises the question of whether the demon mode in $Sr_2RuO_4$ can contribute to pairing interactions of electrons. Since this material has been shown to have demon excitations and superconductivity, it is used as a prototype system to investigate the nature of demons and their role in pairing electrons.

For many years the possibility of having electron excitations contribute to superconductivity has been advanced often. Both excitons [2] and plasma-like excitations [3] have been popular as boson excitations to replace phonons for inducing pairing. A particular type of plasmon-like excitation which we examine here is the so-called "demon" mode which is a neutral acoustic plasmon characterized by out-of-phase oscillations of electrons having different masses or coming from different energy bands. In 1956, Pines called attention to the possible existence of a mode of this type [4]. He coined the name demons based on their characteristically Distinct Electron Motion (DEM). Later references [3] related the name to the screening of heavy "d" electrons by lighter "s,p" electrons. In general, the experimental search for modes of this kind has not been very successful, although some observations have been reported [5]. The precise experiment by Husain *et al.* [1] and careful analysis of their data appear to be convincing proof of the existence of the demon mode.

In examining the role of demons for electron pairing, we follow the approach of Ref. [3] where it was suggested that the mass difference between d- and sp-electrons could produce demon modes and these might serve as a possible pairing mechanism for superconductivity. In some sense the demon mode can be viewed as being similar to an acoustic phonon produced when heavy positive ions are screened by light negative electrons in a jellium model of a metal.

In this study, we begin by determining whether the dynamically screened Coulomb interactions associated with the demon mode can generate attractive pairing interactions. This work relies on an examination of the total dielectric function which when used to screen the Coulomb interaction between elections could give rise to pairing in a BCS-like calculation. Our results indicate that systems of this type or similar systems have promise, however for the specific case of Sr$_2$RuO$_4$, demon-like excitations do not appear to be a successful candidate for a pairing mechanism strong enough to induce superconductivity.

**Methods**

We investigate the superconducting pairing interaction in Sr$_2$RuO$_4$ based on BCS theory and a dielectric-function formalism [6,7] for calculating the electron pairing interaction. The momentum- and energy-dependent BCS gap equation is

$$\Delta_{n\mathbf{k}} = -\frac{1}{2}\sum_{n'\mathbf{k}'} V_{n\mathbf{k},n'\mathbf{k}'} \frac{\Delta_{n'\mathbf{k}'}}{E_{n'\mathbf{k}'}} \tanh\left(\frac{E_{n'\mathbf{k}'}}{2k_BT}\right), \quad (1)$$

where $\Delta_{n\mathbf{k}}$ is the gap function for band $n$ and crystal momentum $\mathbf{k}$, $E_{n\mathbf{k}} = \sqrt{\varepsilon_{n\mathbf{k}}^2 + \Delta_{n\mathbf{k}}^2}$ is the quasiparticle energy, and the pairing interaction $V_{n\mathbf{k},n\mathbf{k}'}$ in general contains both phonon and Coulomb contributions. We will focus on the Coulomb part of the pairing interaction in this study. Taking the average over the momentum, the energy-dependent gap equation becomes

$$\Delta(E) = -\frac{1}{2}\int dE' K^C(E,E') \frac{\Delta(E')}{E'} \tanh\frac{E'}{2k_BT}, \quad (2)$$

where $K^C(E,E')$ is called the Coulomb kernel [8]. We evaluate the Coulomb kernel from the dynamically screened Coulomb matrix elements $W_{n\mathbf{k},n'\mathbf{k}'}(\omega)$

$$K^C(E, E') = N(0) \sum_{n\mathbf{k}n'\mathbf{k}'} W_{n\mathbf{k}n'\mathbf{k}'}(E - E') \frac{\delta(E - \varepsilon_{n\mathbf{k}})}{N(0)} \frac{\delta(E' - \varepsilon_{n'\mathbf{k}'})}{N(0)}, \quad (3)$$

where $N(0)$ is the density of states at the Fermi energy, and $\varepsilon_{n\mathbf{k}}$ is the band structure energy. The Coulomb matrix elements are calculated from the dielectric function $\epsilon(\mathbf{q}, \omega)$

$$W_{n\mathbf{k},n'\mathbf{k}'}(\omega) = \frac{4\pi}{V_{uc}} Re\left(\frac{1}{\epsilon(\mathbf{q}, \omega)}\right) \frac{|\langle u_{n\mathbf{k}}|u_{n'\mathbf{k}'}\rangle|^2}{q^2}, \quad (4)$$

where $V_{uc}$ is the unit cell volume, $\mathbf{q} = \mathbf{k} - \mathbf{k}'$ is the momentum transfer, and $|u_{n\mathbf{k}}\rangle = e^{-i\mathbf{k}\cdot\mathbf{r}}|\psi_{n\mathbf{k}}\rangle$ is the periodic part of the Bloch state. The dielectric function $\epsilon(\mathbf{q}, \omega) = \epsilon_\infty - v(\mathbf{q})\chi^0(\mathbf{q}, \omega)$ is evaluated within the random-phase approximation (RPA), where $\epsilon_\infty$ is the high-frequency dielectric constant, and $v(\mathbf{q}) = 4\pi/q^2$ is the bare Coulomb interaction. We use $\epsilon_\infty = 2.3$ for Sr$_2$RuO$_4$ taken from Ref. [9]. The independent-particle susceptibility $\chi^0(\mathbf{q}, \omega)$ is calculated as

$$\chi^0(\mathbf{q}, \omega) = \frac{2}{N_k V_{uc}} \sum_{nn'\mathbf{k}} (f_{n\mathbf{k}} - f_{n'\mathbf{k}+\mathbf{q}}) \frac{|\langle u_{n\mathbf{k}}|u_{n'\mathbf{k}+\mathbf{q}}\rangle|^2}{\omega + i\eta + \varepsilon_{n\mathbf{k}} - \varepsilon_{n'\mathbf{k}+\mathbf{q}}}, \quad (5)$$

where $N_k$ is the number of $k$ points in the Brillouin Zone, and $f_{n\mathbf{k}}$ is the Fermi-Dirac distribution function.

To calculate the electronic structure of Sr$_2$RuO$_4$, we consider the quasi-2D square lattice of Ru 4$d$ t$_{2g}$ orbitals with lattice parameters a = 3.873 Å and c = 12.7323 Å. The volume per Ru atom is a$^2$c/2. We use the tight-binding Hamiltonian from Ref. [10], which is a 3-band Hamiltonian consisting of 4$d$ t$_{2g}$ orbitals and including the spin-orbit coupling:

$$H_{\sigma,k} = \begin{bmatrix} \varepsilon_k^{yz} - \mu & \varepsilon_k^{off} + i\sigma\lambda & -\sigma\lambda \\ \varepsilon_k^{off} - i\sigma\lambda & \varepsilon_k^{xz} - \mu & i\lambda \\ -\sigma\lambda & -i\lambda & \varepsilon_k^{xy} - \mu \end{bmatrix}, \quad (6)$$

where $\varepsilon_k^{yz} = -2t_2 \cos k_x - 2t_1 \cos k_y$, $\varepsilon_k^{xz} = -2t_1 \cos k_x - 2t_2 \cos k_y$, $\varepsilon_k^{xy} = -2t_3(\cos k_x + \cos k_y) - 4t_4 \cos k_x \cos k_y - 2t_5(\cos 2k_x + \cos 2k_y)$, and $\varepsilon_k^{off} = -4t_6 \sin k_x \sin k_y$. The crystal momenta $k_{x,y}$ are in units of $1/a$. We use the parameters $\lambda = 0.032$ eV, $t_1 = 0.145$ eV, $t_2 = 0.016$ eV, $t_3 = 0.081$ eV, $t_4 = 0.039$ eV, $t_5 = 0.005$ eV, $t_6 = 0.000$ eV, and $\mu = 0.122$ eV.

**Discussion and Results**

Our results for the electronic structure and demon properties are similar to those reported in Reference 1. In Figure 1, the calculated electronic structure of Sr$_2$RuO$_4$ shows that three bands cross the Fermi energy. They are labeled $\beta$, $\alpha$, and $\gamma$, in decreasing order in energy at the $\Gamma$ point. As shown, the density of states (DOS) at the Fermi energy is mostly dominated by the $\beta$ and $\gamma$ bands, and the $\alpha$ band has a prominent peak right above the Fermi energy. The Fermi surface of the $\beta$ band is square-like reflecting its quasi-one-dimensional nature [10]. This feature is crucial in stabilizing the demon mode in this material because it brings a reduction in the phase space for electron-hole excitations thereby prevents the damping of the demon mode, as pointed out in reference [1].

The collective oscillations of electrons in different bands result in the creation of the neutral demon mode. In Sr$_2$RuO$_4$, this mode arises prominently from the out-of-phase motion of electrons in the β and γ bands. The essential information of the demon mode is contained in the inverse dielectric function $\epsilon^{-1}(\mathbf{q}, \omega)$. Figure 2 shows the momentum- and frequency-dependent inverse dielectric function at the low-frequency region where the demon mode is present. As shown in Fig. 2(a), the demon mode appears as a sharp resonance in Im $\epsilon^{-1}(\mathbf{q}, \omega)$, and results in the sign change in the real part of $\epsilon^{-1}(\mathbf{q}, \omega)$ near the demon energy. Fig. 2(b) shows the momentum dependence of Re $\epsilon^{-1}(\mathbf{q}, \omega)$ at $\omega = 80$ meV, which is the typical energy scale of the demon mode in this material. We find that the demon mode results in additional attractive regions near $q_{x,y} \sim 0.1 \frac{2\pi}{a}$. Such attractive regions are present along the (100) and (010) directions because of the anisotropy in the electronic structure. Likewise, the dispersion of the demon mode is sharply defined along the (100) and (010) directions [Figs. 2(c) and 2(d)], but it is overdamped and not seen in the (110) direction [Figs. 2(e) and 2(f)].

We next investigate the Coulomb kernel based on the dielectric function. Figure 3(a) shows the total Coulomb kernel $K^C(0, E)$ and the contributions from each band. We find that the overall magnitude of the Coulomb kernel is mostly below 0.2, which indicates a weak Coulomb repulsion, except for a sharp peak near $E \sim 2$ meV which is attributed to the high density of states coming from $\alpha$ band. Interestingly, the $\gamma$ band has more repulsive Coulomb kernel than the $\beta$ band at the Fermi energy, but the ordering is reversed near 0.1 eV, where the demon mode is strong, the $\gamma$ band kernel becomes even weaker than that of the $\beta$ band. We also note that near $E \sim -0.6$ eV, the Coulomb kernel becomes slightly negative. In our calculations, this is not related to the demon mode, but rather it is a result of the high energy conventional plasmon that disperses downward in energy. However, this may be due to the limitation of the RPA because the experimental measurement of the plasmon dispersion in Sr$_2$RuO$_4$ shows the positive dispersion of the plasmon [11].

Figure 3(b) shows the calculated two-energy Coulomb kernel $K^C(E, E')$. We find no attractive regions except for the plasmon-induced negative parts. The total Coulomb kernel can be decomposed into the momentum-resolved Coulomb kernels $K^C_{n\mathbf{k}}(0, E) = \sum_{n'\mathbf{k}'} W_{n\mathbf{k}n'\mathbf{k}'}(E)\delta(E - \varepsilon_{n'\mathbf{k}'})$, as shown in the Figures 3(c-e) for each band near the demon energy. Although there are some attractive regions in $\beta$ and $\gamma$ band kernels, they do not coincide with the Fermi surface of each band, so the total Coulomb kernel remains positive after the momentum integration. The kernels for the $\beta$ and $\alpha$ bands are mostly constant on their respective Fermi surfaces. However, the $\gamma$ band kernel is anisotropic, showing nearly zero values along (100) and (010) but strongly peaked along the diagonal direction. We suggest that such anisotropy could have implications for the anisotropic superconducting gap of Sr$_2$RuO$_4$ [12].

**Conclusions**

We have examined the dynamically screened Coulomb interaction in Sr$_2$RuO$_4$ with a focus on the role of the demon mode and its contribution to the superconducting pairing interaction. We found the anisotropic nature of the demon mode and identified attractive regions in $(\mathbf{q}, \omega)$ space where the real part of the dielectric function becomes negative. Our main finding is that Sr$_2$RuO$_4$ has an overall weak Coulomb kernel owing to the presence of the demon mode. In particular, the $\gamma$ band kernel is strongly affected by the demon mode, resulting in nearly zero kernels along the (100) and (010) directions. While the demon mode in this material does not produce a total attractive interaction by itself, our results illustrate it can bring significant reduction in the Coulomb repulsion at the relevant energy scale and therefore enhance electron pairing.

**Acknowledgements**

YWC and MLC acknowledge the support of National Science Foundation (NSF) Grant No.DMR-2325410. JI acknowledges the support of the Basic Research Fund of Innonep Inc.

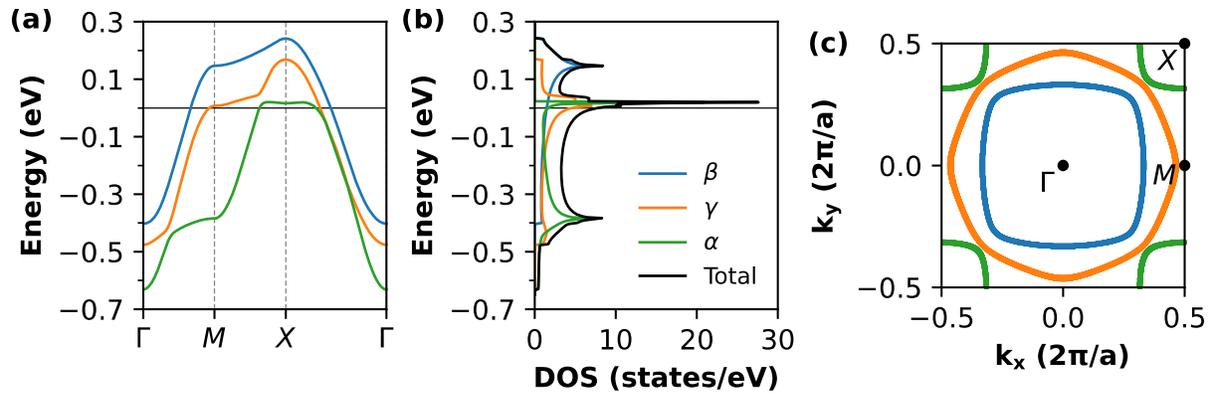

**Figure 1.** Tight-binding electronic structure of $Sr_2RuO_4$ for the one-Ru-cell Brillouin zone. (a) Band structure along the high-symmetry lines, (b) density of states (DOS) per spin, and (c) Fermi surfaces for α (green), β (blue), and γ (orange) bands, respectively.

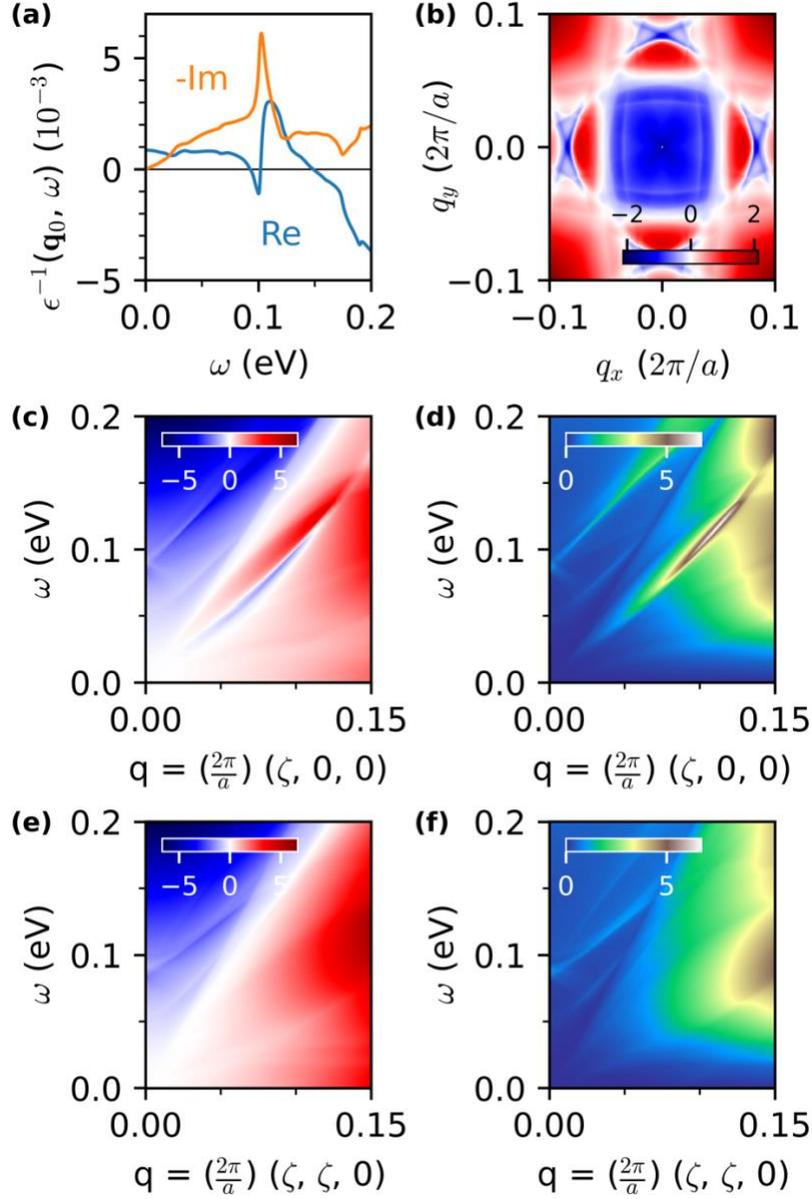

**Figure 2.** Momentum- and frequency-dependent inverse dielectric function $\epsilon^{-1}(\mathbf{q}, \omega)$. (a) Real and imaginary part of the inverse dielectric function as a function of frequency at $\mathbf{q_0} = \frac{2\pi}{a}(0.1, 0, 0)$. The imaginary part shows the resonance for the demon mode, and the real part changes its sign near the demon frequency. (b) Momentum dependence of $Re\ \epsilon^{-1}(\mathbf{q}, \omega)$ at $\omega = 80\ meV$. Because of the demon mode, negative regions are developed near $q \sim 0.1 \frac{2\pi}{a}$ along both $x$ and $y$ directions. (c),(e) Real and (d),(f) imaginary part of $\epsilon^{-1}(\mathbf{q}, \omega)$ along the (100) and (110) directions, respectively. The demon mode is damped away along (110) direction.

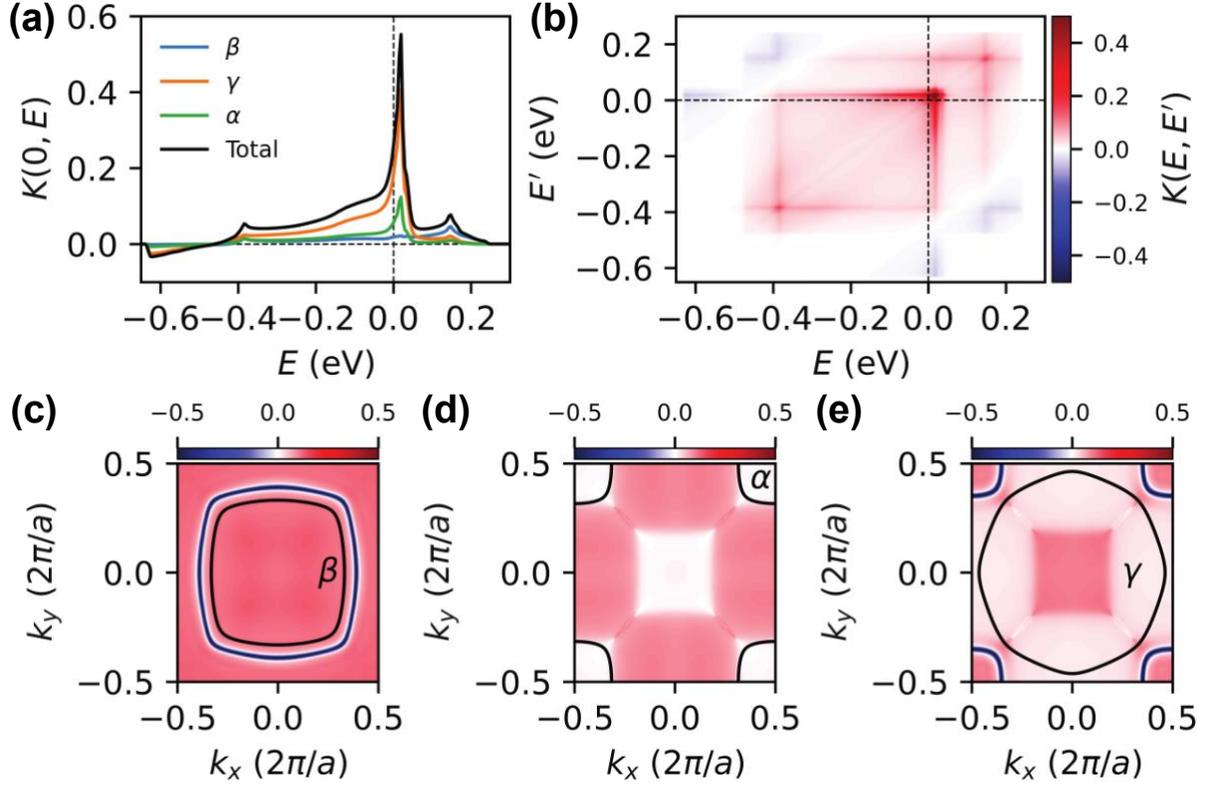

**Figure 3.** (a) Total Coulomb kernel $K^C(0, E)$ as a function of the energy. Blue ($\beta$), orange ($\gamma$), and green ($\alpha$) lines represent the contribution to the total Coulomb kernel from each band. (b) Two-energy Coulomb kernel $K^C(E, E')$ (c-e) Momentum-resolved Coulomb kernels $K^C_{n\mathbf{k}}(0, E)$ at $E = 80\ meV$ for (c) $\beta$, (d) $\alpha$, and (e) $\gamma$ band, respectively. Solid lines represent the Fermi surfaces for each band. Negative parts of $K^C_{n\mathbf{k}}(0, E)$ are not on the Fermi surfaces so the total Coulomb kernel remains repulsive.